\documentclass[12pt]{article}
\usepackage[utf8]{inputenc}
\usepackage{amsmath}
\usepackage{amsfonts}
\usepackage{amssymb} 
\usepackage[final]{feynmp}
\usepackage{caption}
\usepackage{cite}
\usepackage{graphicx}
\usepackage[font={footnotesize}]{caption}
\usepackage{color}
\usepackage[colorlinks=true
,urlcolor=blue
,citecolor=blue
,linkcolor=blue
,pagecolor=blue
,linktocpage=true
,pdfproducer=medialab
]{hyperref}
\usepackage[a4paper,width=15.2cm]{geometry}
\makeatletter \renewcommand{\@dotsep}{10000} \makeatother
\newcommand\prd[3]{{\it Phys.\ Rev.\  \bf D }{\bf #1} (#2) #3}

\newcommand\plb[3]{{\it Phys.\ Lett.\  \bf B }{\bf #1} (#2) #3}
\newcommand\npb[3]{{\it Nucl.\ Phys.\  \bf B }{\bf #1} (#2) #3}
\newcommand{\vev}[1]{\left<#1\right>}
\begin{document}

\begin{flushright}
\end{flushright}

\vspace{1cm}
\begin{center}
{\Large\bf  Masses of Third Family Vector-like Quarks and Leptons in Yukawa-Unified $E_6$
 } \vspace{1cm}

{\large  Aditya~Hebbar$^{a,}$\footnote{E-mail: adityah@udel.edu }, George~K.~Leontaris$^{b,}$\footnote{E-mail: leonta@uoi.gr
 } and   Qaisar~Shafi$^{c,}$\footnote{E-mail: shafi@bartol.udel.edu
} } \vspace{.5cm}

{ \it
$^a$Department of Physics and Astronomy, \\
University of Delaware, Newark, DE 19716, USA
} \vspace{.5cm}

{ \it
$^b$Physics Department, Theory Division, \\
University of Ioaninna, GR-45110 Ioannina, Greece
} \vspace{.5cm}

{ \it
$^c$Bartol Research Institute, Department of Physics and Astronomy, \\
University of Delaware, Newark, DE 19716, USA
} \vspace{.5cm}

\vspace{1.5cm}
 {\bf Abstract}
\end{center}

\noindent In supersymmetric $E_6$ the masses of the third family quarks and charged lepton, t-b-$\tau$, as well as the masses of the vector-like quarks and leptons, D-$\bar{\text{D}}$ and L-$\bar{\text{L}}$, may arise from the coupling $27_3$ x $27_3$ x $27_H$, where $27_3$ and $27_H$ denote the third family matter and Higgs multiplets respectively. We assume that the SO(10) singlet component in $27_H$ acquires a TeV scale VEV which spontaneously breaks U(1)$_\psi$ and provides masses to the vector-like particles in $27_3$, while the MSSM doublets in $27_H$ provide masses to t, b and $\tau$. Imposing Yukawa coupling unification $h_t=h_b=h_{\tau}=h_D=h_L$ at $M_{GUT}$ and employing the ATLAS and CMS constraints on the $Z^\prime_\psi$ boson mass, we estimate the lower bounds on the third family vector-like particles D-$\bar{\text{D}}$ and L-$\bar{\text{L}}$ masses to be around 5.85 TeV and 2.9 TeV respectively.
These bounds apply in the supersymmetric limit.

\newpage


\section*{Introduction}

Yukawa coupling unification (YU) $h_t=h_b=h_\tau$ of third family charged fermions \cite{anant} has  received a great deal of attention \cite{hall} \cite{other}. While originally implemented in SO(10) grand unification \cite{so10}, YU has also been studied \cite{recent} in the framework of SU(4)$_c$ x SU(2)$_L$ x SU(2)$_R$ (4-2-2) gauge group \cite{pati} with some interesting results. For instance, in 4-2-2 models assuming plausible supersymmetry (SUSY) breaking scenarios such as non universal Higgs and scalar masses (NUHM2), one finds that a scenario consistent with the observed (LSP) dark matter abundance and various other experiments requires a NLSP gluino. This is certainly being tested at the LHC. 
In a different scenario also with t-b-$\tau$ YU but non-universal gaugino masses \cite{nugm}, the colored sparticles lie in the multi-TeV range, while the scalar partners of the leptons are significantly lighter with some also contributing in an essential way to the muon anomalous magnetic moment \cite{g-2}.\\

\noindent In this letter we propose a new extended version of t-b-$\tau$ Yukawa unification inspired by $E_6$ grand unification \cite{gursey}\cite{achimon}\cite{shafi}\cite{recentE6}. In contrast to SO(10) where t-b-$\tau$ YU only holds in a supersymmetric model, the proposed extended YU, which includes t-b-$\tau$ YU, can also be realized in non-supersymmetric $E_6$. In this paper we will restrict our attention to third family extended YU in supersymmetric $E_6$. Assuming that the gauge symmetry U(1)$_\psi$ in $E_6$ is spontaneously broken in the TeV region, we will exploit the extended YU boundary conditions to infer the lower bounds on the masses of the third family vector-like quarks and lepton doublets predicted in $E_6$.\\

\noindent The discussion in the paper proceeds as follows. After a brief overview of t-b-$\tau$ YU, we present some figures based on one-loop RGEs that implement this scenario in the framework of MSSM, as well as MSSM plus three ($5 + \bar{5}$) multiplets which are present in $E_6$ models. The one-loop threshold corrections \cite{hall}\cite{pierce} play an important role here. We estimate the unified Yukawa coupling at $M_{GUT}$ to be approximately 0.6 in MSSM (Figure \ref{YU_mssm}), in agreement with previous work \cite{yupointsix}. We then consider MSSM plus three ($5+\bar{5}$) multiplets and show a plot displaying gauge coupling unification (Figure \ref{vlikeGU}). 	We estimate the unified Yukawa coupling at $M_{GUT}$ in the presence of three ($5+\bar{5}$) fields to be 0.35 (Figure \ref{YU_vlike}).  To our knowledge, t-b-$\tau$ YU with three additional vector-like families has not been previously discussed. Finally, we discuss extended Yukawa coupling unification motivated by the $E_6$ invariant Yukawa coupling 27$_3$ x 27$_3$ x 27$_H$, where the subscripts denote third family matter fields and the Higgs 27-plet. With the assumption that the MSSM Higgs doublets H$_u$, H$_d$ and the SO(10) singlet field N that breaks U(1)$_\psi$ arise from this 27$_H$, we are led to extended YU. Namely, at M$_{GUT}$, $h_t = h_b = h_\tau = h_D = h_L$, where $h_D$ and $h_L$ respectively refer to the Yukawa couplings  to N of the vector-like color triplets and SU(2) doublets. Employing the lower bound on $Z^\prime_\psi$ boson mass provided by ATLAS/CMS \cite{zprime}, we estimate the lower bound on third family D and L masses to be 5.85 TeV and 2.9 TeV respectively. In the $E_6$ case the Yukawa coupling at $M_{GUT}$ is estimated to be around 0.3-0.35. In the summary section we note that this value is intriguingly close to the estimates of 0.3-0.5 for the unified third family Yukawa coupling obtained in certain F-theory models based on the $E_8$ gauge symmetry.

\section*{t-b-$\tau$ Yukawa Unification }

In SUSY SO(10) it is plausible that the third family charged fermions in the 16-plet primarily acquire masses from the invariant coupling 16$_3$ x 16$_3$ x 10$_H$, with the MSSM doublets contained in $10_H$. This yields $h_t = h_b = h_\tau$ at M$_{GUT}$ \cite{anant}. Note that SUSY plays an essential role here and without it, t-b-$\tau$ YU would not be possible in SO(10) barring additional assumptions. The one loop renormalization group equations for the Yukawa couplings are \cite{Falck:1985aa,Barger:1992ac,Barger:1993gh,Martin:1993zk}:
\begin{eqnarray}
\frac{dh_t}{dt}&=&\frac{h_t}{16\pi^2}\left(6h_t^2+h_b^2-\left(\frac{16}{3}g_3^2+3g_2^2+\frac{13}{15}g_1^2\right)\right),\nonumber\\
\frac{dh_b}{dt}&=&\frac{h_b}{16\pi^2}\left(6h_b^2+h_t^2+h_\tau^2-\left(\frac{16}{3}g_3^2+3g_2^2+\frac{7}{15}g_1^2\right)\right),\nonumber\\
\frac{dh_\tau}{dt}&=&\frac{h_\tau}{16\pi^2}\left(3h_b^2+4h_\tau^2-\left(3g_2^2+\frac{9}{5}g_1^2\right)\right),
\label{mssm}
\end{eqnarray}
with t = log(Q), where Q is the renormalization scale.\\

\begin{figure}[!h]
\begin{minipage}{0.5\textwidth}
\centering
\includegraphics[width=0.9\linewidth, height=0.7\linewidth]{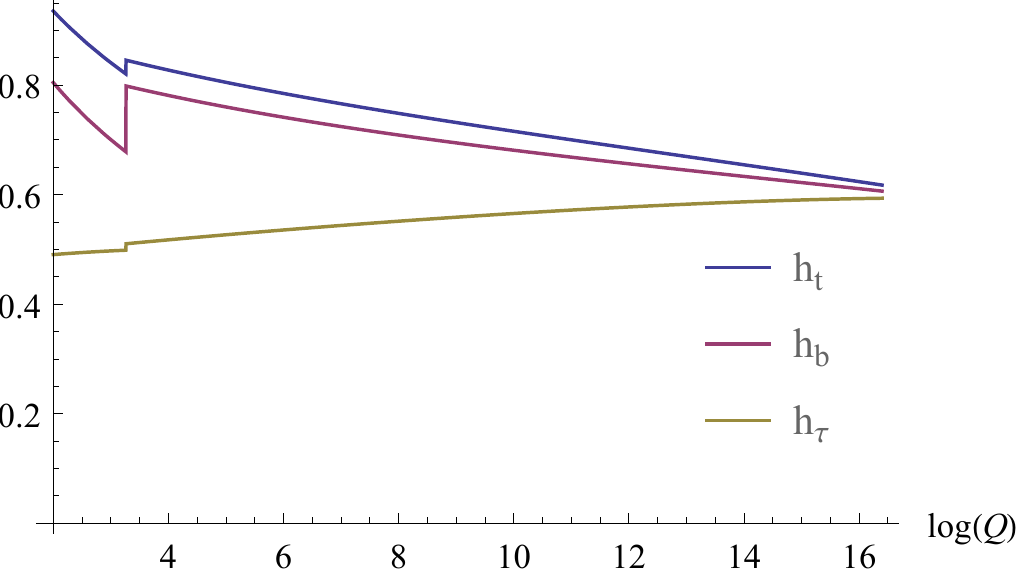}
\end{minipage}%
\begin{minipage}{0.5\textwidth}
\centering
\includegraphics[width=0.9\linewidth, height=0.7\linewidth]{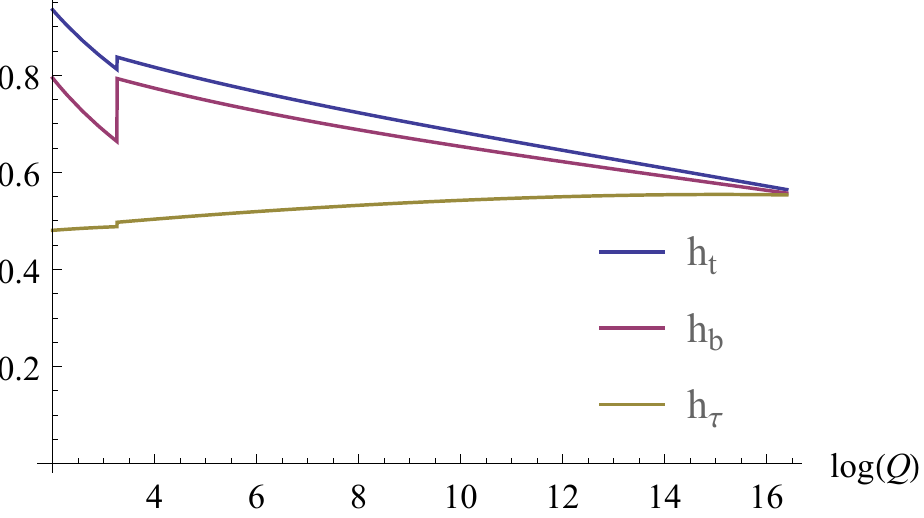}
\end{minipage}
\caption{\footnotesize Yukawa coupling unification in MSSM  with tan $\beta$ = 50 (left) and tan $\beta$ = 51 (right). }
\label{YU_mssm}
\end{figure}

\noindent Using plausible values for the masses of SUSY particles ($m_{\tilde t_1}$, $m_{\tilde t_2}$, $m_{\tilde b_1}$, $m_{\tilde b_2}$ and $m_{\tilde g}$ in the range 2-3 TeV), $\mu$ parameter ($\approx$ 0.5-0.7 TeV) and $A_t \approx 2.5$ TeV, we present two plots (Figure \ref{YU_mssm}) showing how t-b-$\tau$ YU can be realized. Note that in this paper we do not specify any particular SUSY breaking scenario, and these parameters will be employed throughout the paper. The MSSM parameter tan $\beta$ is around 50, and we confirm previous results that the unified Yukawa coupling at M$_{GUT}$ is approximately 0.6\cite{yupointsix}. \\

\noindent In the evolution of RGEs we take into account one loop threshold corrections due to sparticle loops\cite{hall}\cite{pierce}. In practice, the largest correction is often to the bottom Yukawa coupling. An approximate expression for the bottom correction (in the limit of the masses of gluino and the top and bottom squarks being approximately equal) is given by:
\begin{align}
\delta h_b^{\rm finite}\approx \frac{g_3^2}{12\pi^2}\frac{\mu m_{\tilde g}
\tan\beta}{m_{\tilde b}^2}+
                         \frac{h_t^2}{32\pi^2}\frac{\mu A_t \tan\beta}{m_{\tilde t}^2},
\end{align}\\
where m$_{\tilde b} \approx \frac{m_{\tilde b_1}+m_{\tilde b_2}}{2}$ and m$_{\tilde t} \approx \frac{m_{\tilde t_1}+m_{\tilde t_2}}{2}$ denote the average bottom and top squark masses respectively. \\

%
\section*{Gauge and t-b-$\tau$ Yukawa unification in MSSM and three $(5 + \bar{5})$ multiplets}

\noindent We now consider MSSM plus three $(5 + \bar{5})$ multiplets. The one loop RGEs for the gauge couplings at the 1-loop level are given by:
\begin{eqnarray}
\alpha^{-1}_i(Q) = \alpha^{-1}_i(Q_0)- \frac{b_i}{2\pi}\text{log}\left(\frac{Q}{Q_0}\right).
\end{eqnarray}
\noindent The coefficients $b_i$ depend on the matter content of the theory. For a SUSY theory with $n_g$ families, $n_H$ Higgs doublets and $n_v$ vector-like families, the coefficients $b_i$ are given as follows~\cite{Falck:1985aa,mac,Cvetic:1998uw,Barger:1992ac,Barger:1993gh,Martin:1993zk}:

\begin{eqnarray}
b_1&=&0+2n_g+\frac{3}{10}n_H+n_v,\\
b_2&=&-6+2n_g+\frac{1}{2}n_H+n_v,\\
b_3&=&-9+2n_g+n_v.
\end{eqnarray}

\begin{figure}[!htb]
\centering
\includegraphics[width=0.7\textwidth, height=0.4\textwidth]{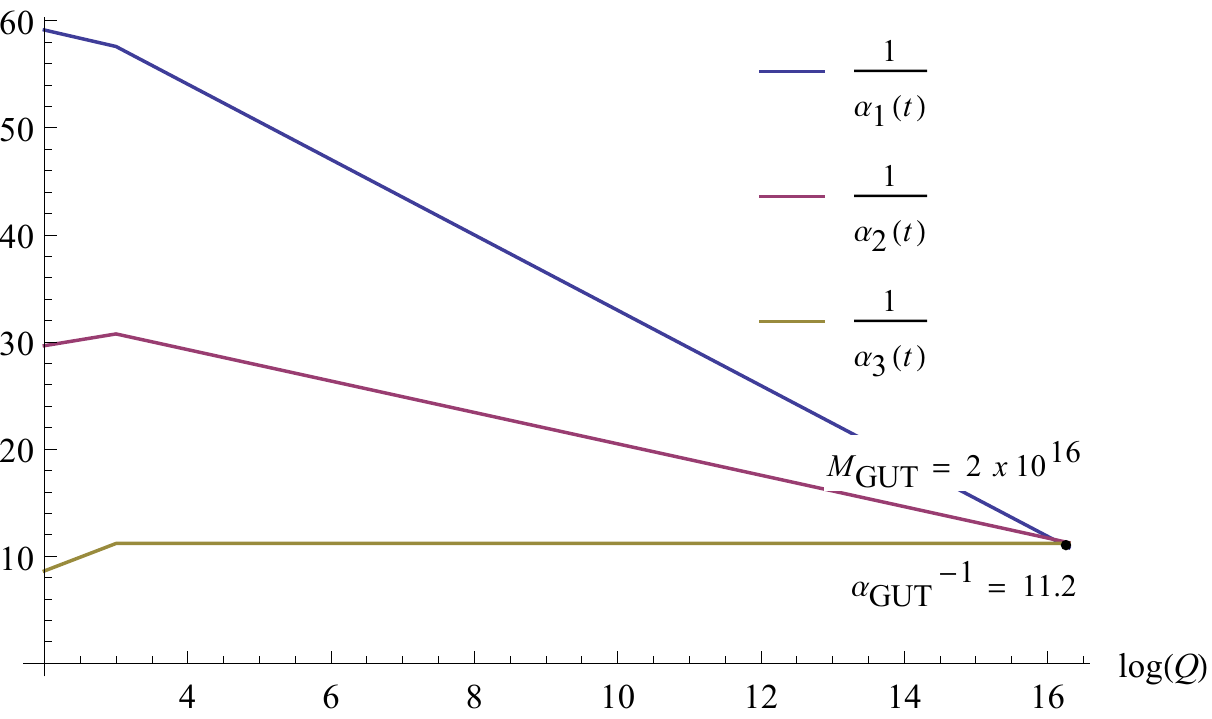}
\caption{\small Gauge coupling unification in MSSM and three TeV scale ($5+\bar{5}$) multiplets. \\ For the $E_6$ model, the gauge coupling g$_{\psi}$ associated with the extra U(1)$_\psi$ symmetry is approximately equal to g$_1$ of the MSSM in between TeV scale and the M$_{GUT}$ scale. }
\label{vlikeGU}
\end{figure}

\noindent For MSSM and three vector-like ($5+\bar{5}$) multiplets, we get the coefficients: 
\begin{equation}\left(\frac{48}{5}, 4, 0\right)\end{equation}
\noindent The presence of these new particles increases the value of the unified gauge coupling strength at $M_{GUT}$ as shown in Figure \ref{vlikeGU}.

\begin{figure}[!h]
\begin{minipage}{0.5\textwidth}
\centering
\includegraphics[width=0.9\linewidth, height=0.7\linewidth]{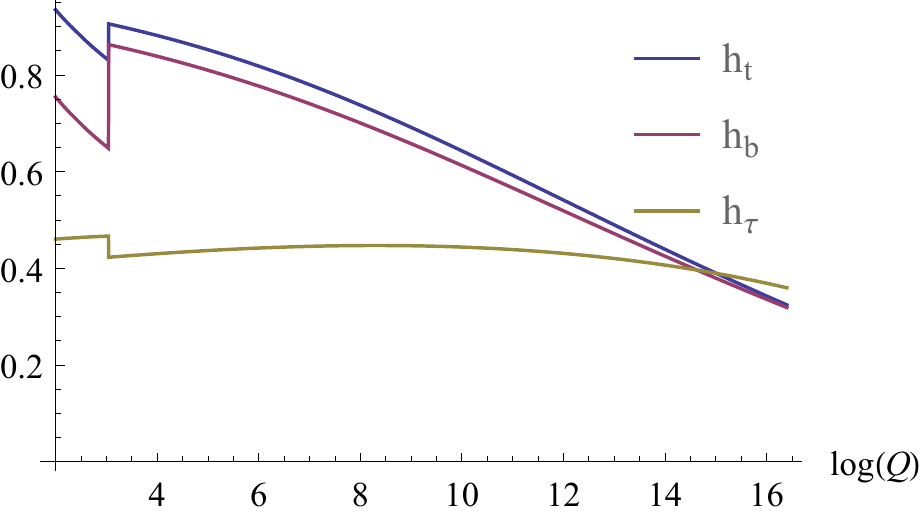}
\caption*{tan $\beta$ = 48}
\end{minipage}%
\begin{minipage}{0.5\textwidth}
\centering
\includegraphics[width=0.9\linewidth, height=0.7\linewidth]{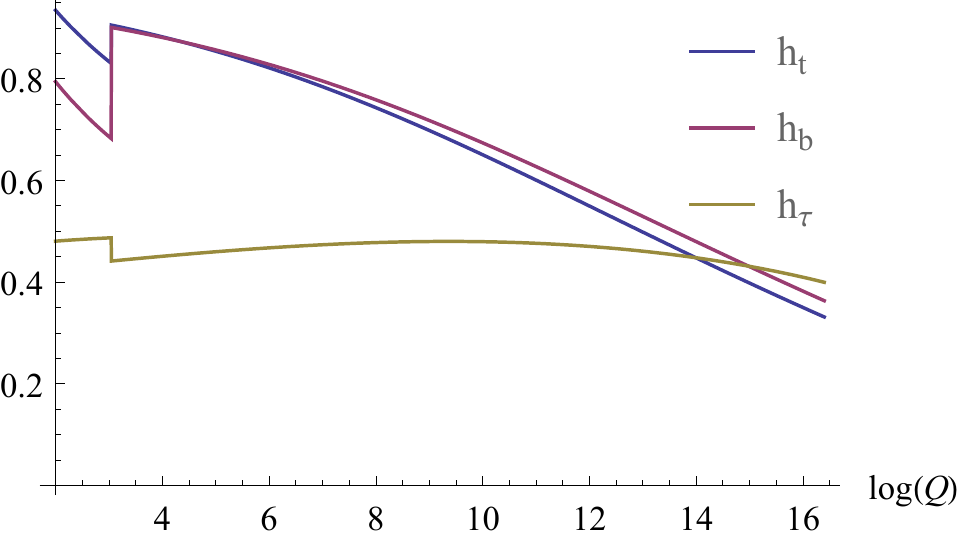}
\caption*{tan $\beta$ = 50}
\end{minipage}
\caption{\footnotesize Yukawa coupling unification in MSSM and three TeV scale ($5+\bar{5}$) multiplets. }
\label{YU_vlike}
\end{figure}

\noindent Next we implement t-b-$\tau$ YU in MSSM plus three families of TeV scale ($5 + \bar{5}$) particles. The RGEs for the Yukawa couplings remain the same as that for MSSM (\ref{mssm}). The evolution of the Yukawa couplings in this case is displayed in Figure \ref{YU_vlike}. As one might expect, the larger gauge couplings have a somewhat greater impact on the Yukawa couplings, and the Yukawa coupling at M$_{GUT}$ is estimated to be around 0.35.\\

\section*{$E_6$ Yukawa unification }

\noindent We next proceed to E$_6$ grand unification and make the simplifying assumption that the Higgs scalars that spontaneously break the MSSM gauge symmetry and U(1)$_\psi$ arise from the same Higgs 27-plet (27$_H$). \noindent Under the decomposition $E_6$ $\rightarrow$ SO(10) x U(1)$_\psi$, the 27-plet contains the following fields: 
\begin{equation}
27 \rightarrow 16_{1} + 10_{-2} + 1_{4},
\end{equation}
\noindent where the subscripts denote $2\sqrt{6}$ Q$_\psi$, with Q$_\psi$ being the normalized U(1)$_\psi$ charge  \cite{lw}\cite{leike}. \\

\noindent The Yukawa couplings D$\bar{\text{D}}$N and L$\bar{\text{L}}$N  provide masses to these particles, while the coupling H$_u$H$_d$N can yield the MSSM $\mu$ term. Indeed, this latter feature provides a good motivation for breaking U(1)$_\psi$ at the TeV scale\cite{lw}. From the third family Yukawa coupling 27$_3$ x 27$_3$ x 27$_H $, we obtain the asymptotic YU relation $h_t = h_b = h_\tau = h_D = h_L$. The one loop RGEs for Yukawa coupings are given by \cite{Falck:1985aa}\cite{lw}:

\begin{eqnarray}
\frac{dh_t}{dt}&=&\frac{h_t}{16\pi^2}\left(6h_t^2+h_b^2-\left(\frac{16}{3}g_3^2+3g_2^2+\frac{13}{15}g_1^2+\frac{1}{2}g_{\psi}^2\right)\right),\nonumber\\
\frac{dh_b}{dt}&=&\frac{h_b}{16\pi^2}\left(6h_b^2+h_t^2+h_\tau^2-\left(\frac{16}{3}g_3^2+3g_2^2+\frac{7}{15}g_1^2+\frac{1}{2}g_{\psi}^2\right)\right),\nonumber\\
\frac{dh_\tau.}{dt}&=&\frac{h_\tau}{16\pi^2}\left(3h_b^2+4h_\tau^2-\left(3g_2^2+\frac{9}{5}g_1^2+\frac{1}{2}g_{\psi}^2\right)\right),\nonumber\\
\frac{dh_L}{dt}&=&\frac{h_L}{16\pi^2}\left(4h_L^2+3h_D^2-\left(3g_2^2+\frac{3}{5}g_1^2+2g_{\psi}^2\right)\right),\nonumber\\
\frac{dh_D}{dt}&=&\frac{h_D}{16\pi^2}\left(5h_D^2+2h_L^2-\left(\frac{16}{3}g_3^2+\frac{4}{15}g_1^2+2g_{\psi}^2\right)\right),
\end{eqnarray}
with t = log(Q), where Q is the renormalization scale.\\

\noindent In the evolution of MSSM gauge couplings we assume that all three vector-like families have masses in the TeV range. However, Yukawa unification as presented above only applies to $27_3$, the third family. We assume that there is negligible mixing between the $\bar{5}$ multiplets contained in the 16-plet and 10-plet of SO(10). We also assume that the $16_H$ component in the $27_H$ is decoupled from low energy physics, through appropriate fine-tuning.\\

\noindent We explore a few scenarios with varying sparticles masses and A terms to convince ourselves that the third family unified Yukawa coupling at $M_{GUT}$ is around 0.35.\\

\begin{figure}[!htb]
\begin{minipage}{0.5\textwidth}
\centering
\includegraphics[width=0.9\linewidth, height=0.7\linewidth]{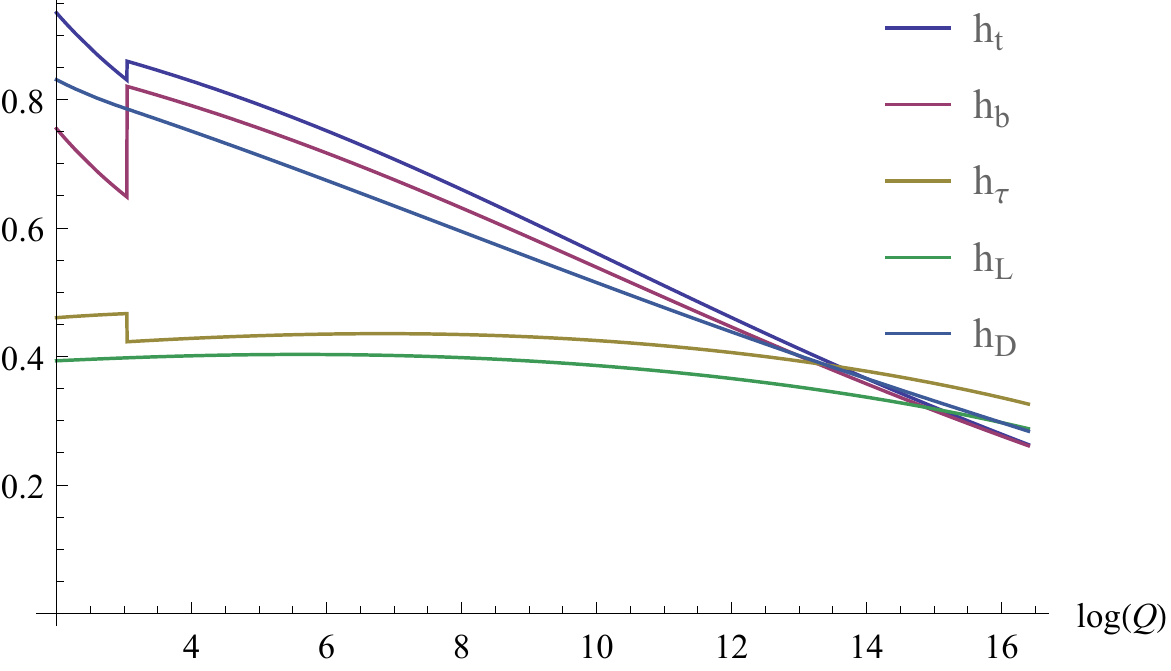}
\end{minipage}%
\begin{minipage}{0.5\textwidth}
\centering
\includegraphics[width=0.9\linewidth, height=0.7\linewidth]{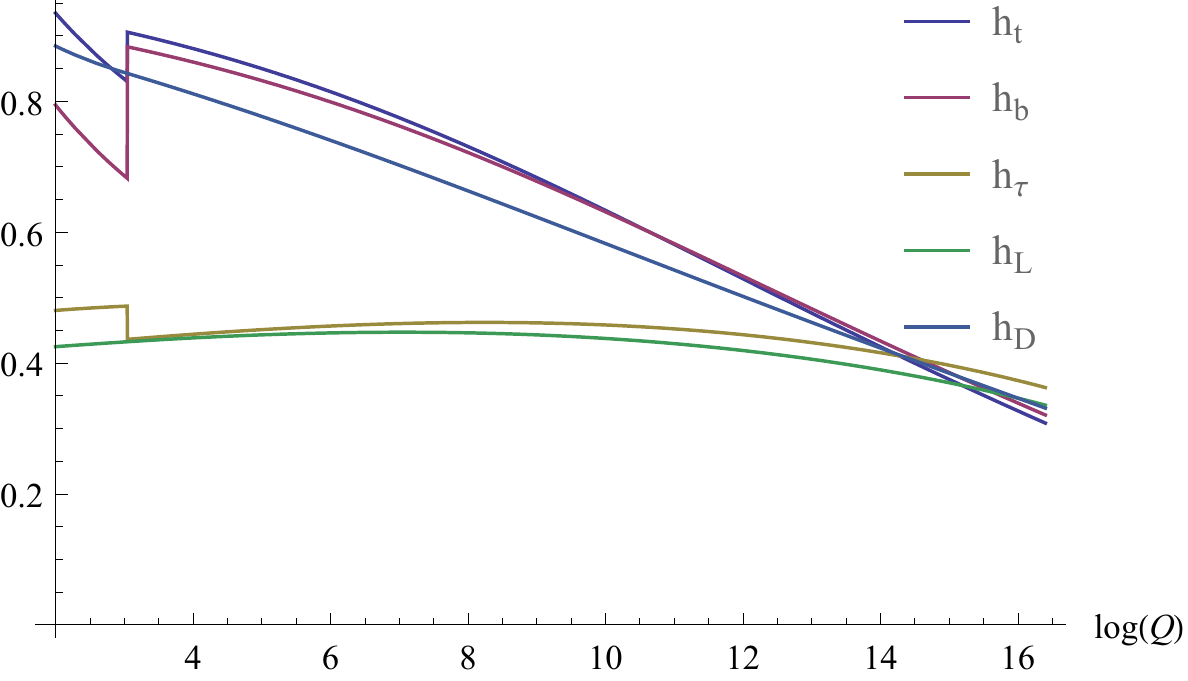}

\end{minipage}
\caption{\footnotesize Yukawa coupling unification in $E_6$ with tan $\beta$ = 48 (left) and tan $\beta$ = 50 (right). }
\label{YU_E6}
\end{figure}

\noindent Assuming gauge unification, we set the U(1)$_\psi$ gauge coupling $g_\psi$ equal to MSSM U(1) gauge coupling $g_1$ at $M_{GUT}$. With $b_\psi$ = Tr $Q_\psi^2$ = 9.66\cite{lw} being approximately the same as $b_1$ = 9.6 for $g_1$, the two couplings stay close to each other between $M_{GUT}$ and the TeV scale.\\

\noindent From gauge coupling unification (Figure \ref{vlikeGU}) we estimate that the U(1)$_\psi$ gauge coupling g$_{\psi} \approx$ g$_1$ = 0.47 at LHC energies. Together with the lower bound on the $Z^\prime_\psi$ boson mass of 2.79 TeV \cite{zprime}, the VEV of the SO(10) singlet scalar field N is estimated as follows:
\begin{eqnarray}
 m_{Z^\prime_\psi}&=& \frac{4}{2\sqrt{6}}g_{\psi} \vev{N},\nonumber\\
\implies \vev{N} &>& 7.3 \text{ TeV}.
\end{eqnarray}

\noindent Combining this with h$_D$ and h$_L$ evaluated in the TeV
range  (Figure \ref{YU_E6}), we estimate that the lower 
bounds on the masses of the third family
 vector-like D and L fields are
\begin{eqnarray} 
 M_D &\gtrsim& 5.85\,{\rm TeV }\\
 M_L&\gtrsim &2.90\, {\rm TeV } .
 \end{eqnarray}
 Clearly these bounds do not take into account supersymmetry breaking.

\section*{Summary}

\noindent We have extended the idea of t-b-$\tau$ Yukawa unification in supersymmetric SO(10) by including the Yukawa couplings of vector-like fields that appear in the matter 27-plet of $E_6$. Requiring unification of these Yukawa couplings at $M_{GUT}$, and taking into account both gauge coupling unification and the ATLAS/CMS lower bound on the appropriate $Z^\prime$ boson mass enables us to estimate lower bounds on the third family color triplet and SU(2) doublet vector-like fields. We should remark here that the first two families of vector-like quarks and leptons may well be considerably lighter than the third family, in which case some of them may be accessible at the LHC. It is interesting to note that certain F-theory constructions utilizing $E_8$ and $E_6$ gauge symmetries predict unified Yukawa coupling $\sim$ 0.3-0.5 for the third family fields including the vector-like ones contained in $E_6$ \cite{Cecotti:2009zf}. This is intriguingly close to the value of 0.3-0.35 that we estimate for extended YU in the $E_6$ model and deserves further exploration.

\section*{Acknowledgements}

We thank Tianjun Li, Nobuchika Okada and especially Cem Salih Un for helpful discussions. Q.S. is supported in part by the DOE grant DOE-SC-0013880.

\end{document}